\documentclass[runningheads]{llncs}
\usepackage{graphicx}
\usepackage{array,multirow}
\usepackage{booktabs}       
\usepackage{dirtytalk}
\usepackage{amsfonts}       
\usepackage{nicefrac}       
\usepackage{microtype}      
\usepackage{amsmath}
\usepackage{amssymb}
\usepackage[utf8]{inputenc} 
\usepackage[T1]{fontenc}    
\usepackage{hyperref}       
\usepackage{url}            
\usepackage{lipsum}
\usepackage{lmodern}
\usepackage{mathtools}
\usepackage{color}

\begin{document}
%
\title{Extended Graph Assessment Metrics for Regression and Weighted Graphs}
\author{Tamara T. Mueller\inst{1} \and
Sophie Starck\inst{1} \and
Leonhard F. Feiner\inst{1,2} \and
Kyriaki-Margarita Bintsi\inst{3} \and
Daniel Rueckert\inst{1,3} \and
Georgios Kaissis\inst{1,2,4} 
}

\authorrunning{Mueller et al.}
%
\institute{Institute for AI in Medicine and Healthcare, Faculty of Informatics, Technical University of Munich \and
Department of Diagnostic and Interventional Radiology, Faculty of Medicine, Technical University of Munich \and
BioMedIA, Department of Computing, Imperial College London \and
Institute for Machine Learning in Biomedical Imaging, Helmholtz-Zentrum Munich \\
\email{tamara.mueller@tum.de}
}

\maketitle              
\begin{abstract}
When re-structuring patient cohorts into so-called population graphs, initially independent patients can be incorporated into one interconnected graph structure. 
This population graph can then be used for medical downstream tasks using graph neural networks (GNNs). 
The construction of a suitable graph structure is a challenging step in the learning pipeline that can have severe impact on model performance. 
To this end, different graph assessment metrics have been introduced to evaluate graph structures. 
However, these metrics are limited to classification tasks and discrete adjacency matrices, only covering a small subset of real-world applications. 
In this work, we introduce extended graph assessment metrics (GAMs) for regression tasks and weighted graphs. 
We focus on two GAMs in specific: \textit{homophily} and \textit{cross-class neighbourhood similarity} (CCNS). We extend the notion of GAMs to more than one hop, define homophily for regression tasks, as well as continuous adjacency matrices, and propose a light-weight CCNS distance for discrete and continuous adjacency matrices. 
We show the correlation of these metrics with model performance on different medical population graphs and under different learning settings, using the TADPOLE and UKBB datasets. 
\end{abstract}

\section{Introduction}
The performance of graph neural networks can be highly dependent on the graph structure they are trained on \cite{ma2021homophily,luan2021heterophily}. 
To this end, several graph assessment metrics (GAMs) have been introduced to evaluate graph structures and shown strong correlations between specific graph structures and the performance of graph neural networks (GNNs) \cite{luan2022we,ma2021homophily,luan2021heterophily}. 
Especially in settings, where the graph structure is not provided by the dataset but needs to be constructed from the data, GAMs are the only way to assess the quality of the constructed graph. 
This is for example the case when utilising so-called population graphs on medical datasets. 
Recent works have furthermore shown that learning the graph structure in an end-to-end manner, can improve performance on population graphs \cite{kazi2022differentiable}. 
Some of these methods that learn the graph structure during model training operate with fully connected, weighted graphs, where all nodes are connected with each other and the tightness of the connection is determined by a learnable edge weight. 
This leads to a different representation of the graph, which does not fit the to-date formulations of GAMs. Furthermore, existing metrics are tailored to classification tasks and cannot be easily transformed for equally important regression tasks. 
The contributions of this work are the following: (1) We extend existing metrics to allow for an assessment of multi-hop neighbourhoods. (2)  We introduce an extension of the homophily metric for regression tasks and continuous adjacency matrices and (3) define a cross-class neighbourhood similarity (CCNS) distance metric and extend CCNS to learning tasks that operate on continuous adjacency matrices. Finally, (4) we show these metrics' correlation to model performance on different medical and synthetic datasets.
The metrics introduced in this work can find versatile applications in the area of graph deep learning in medical and non-medical settings, since they strongly correlate with model performance and give insights into the graph structure in various learning settings. 

\section{Background and Related Work}
\subsection{Definition of graphs}
A discrete graph $\mathit{G}:=(\mathit{V}, \mathit{E})$ is defined by a set of $n$ nodes $\mathit{V}$ and a set of edges $\mathit{E}$, connecting pairs of nodes. The edges are unweighted and can be represented by an adjacency matrix $\mathbf{A}$ of shape $n \times n$, where $\mathbf{A}_{ij} = 1$ if and only if $e_{ij} \in \mathit{E}$ and $0$ otherwise.
A continuous/weighted graph $\mathit{G}_w:=(\mathit{V}_w, \mathit{E}_w, \mathbf{W})$, assigns a (continuous) weight to every edge in $\mathit{E}_w$, summarised in the weight matrix $\mathbf{W}$. Continuous graphs are for example required in cases where the adjacency matrix is learned in an end-to-end manner and backpropagation through the adjacency matrix needs to be feasible.
A neighbourhood $\mathcal{N}_v$ of a node $v$ contains all direct neighbours of $v$ and can be extended to $k$ hops by $\mathcal{N}^{(k)}_v$.
For this work, we assume familiarity with GNNs \cite{bronstein2017geometric}.

\subsection{Homophily}
\label{sec:background_homophily}
Homophily is a frequently used metric to assess a graph structure that is correlated to GNN performance \cite{luan2021heterophily}. It quantifies how many neighbouring nodes share the same label \cite{luan2021heterophily} as the node of interest. There exist three different notions of homophily: edge homophily \cite{kim2022find}, node homophily \cite{pei2020geom}, and class homophily \cite{lim2021new,luan2021heterophily}. 
Throughout this work, we use node homophily, sometimes omitting the term \say{node}, only referring to \say{homophily}.

\begin{definition}[Node homophily]
  Let $\mathit{G} := (\mathit{V}, \mathit{E})$ be a graph with a set of node labels $\mathit{Y} := \{y_u; u \in \mathit{V}\}$ and $\mathcal{N}_v$ be the set of neighbouring nodes to node $v$. Then $\mathit{G}$ has the following node homophily:

\begin{equation}
  h(\mathit{G}, \mathit{Y}) := \frac{1}{\vert \mathit{V} \vert} \sum_{v \in \mathit{V}}\frac{\left\vert \{ u | u \in \mathcal{N}_v, Y_u = Y_v \} \right\vert}{\vert \mathcal{N}_v \vert},
  \label{eq:homophily}
\end{equation} 
where $\vert \cdot \vert$ indicates the cardinality of a set.
\end{definition}
A graph $G$ with node labels $Y$ is called \textit{homophilous}/\textit{homophilic} when $h(G, Y)$ is large (typically larger than 0.5) and \textit{heterophilous}/\textit{heterophilic} otherwise \cite{kim2022find}. 
 
\subsection{Cross-class neighbourhood similarity}
Ma et al. \cite{ma2021homophily} introduce a metric to assess the graph structure for graph deep learning, called cross-class neighbourhood similarity (CCNS). This metric indicates how similar the neighbourhoods of nodes with the same labels are over the whole graph -- irrespective of the labels of the neighbouring nodes.

\begin{definition}[Cross-class neighbourhood similarity]
Let $\mathit{G} = (\mathit{V}, \mathit{E})$, $\mathcal{N}_v$, and $Y$ be defined as above. Let $C$ be the set of node label classes, and $\mathcal{V}_c$ the set of nodes of class $c$. Then the CCNS of two classes $c$ and $c'$ is defined as follows:

\begin{equation}
    \operatorname{CCNS}(c, c') = \frac{1}{\vert \mathcal{V}_c \vert \vert \mathcal{V}_{c'} \vert} \sum_{u \in \mathcal{V}, v \in \mathcal{V'}}{\operatorname{cossim}(d(u),d(v))}.
\end{equation}
$d(v)$ is the histogram of a node $v$'s neighbours' labels and $\operatorname{cossim}(\cdot,\cdot)$ the cosine similarity. 
\end{definition}

\section{Extended Graph Metrics}
In this section, we introduce our main contributions by defining new extended GAMs for regression tasks and continuous adjacency matrices. We propose (1) a unidimensional version of CCNS which we call \textit{CCNS distance}, which is easier to evaluate than the whole original CCNS matrix, (2) an extension of existing metrics to $k$-hops, (3) GAMs for continuous adjacency matrices, and (4) homophily for regression tasks.

\subsection{CCNS distance}
The CCNS of a dataset with $n$ classes is an $n \times n$ matrix, which can be large and cumbersome to evaluate. The most desirable CCNS for graph learning has high intra-class and low inter-class values, indicating similar neighbourhoods for the same class and different neighbourhoods between classes.
We propose to collapse the CCNS matrix into a single value by evaluating the $L_1$ distance between the CCNS and the identity matrix, which we term \textit{CCNS distance}. 

\begin{definition}[CCNS distance]
Let $\mathit{G} = (\mathit{V}, \mathit{E})$, $C$, CCNS be defined as above. Then the CCNS distance of $\mathit{G}$ is defined as follows:
\begin{equation}
    D_{\operatorname{CCNS}} := \frac{1}{n} \sum{\lVert \operatorname{CCNS} - \mathbb{I} \rVert_1 },
\end{equation}
where $\mathbb{I}$ indicates the identity matrix and $\lVert \cdot \rVert_1$ the $L_1$ norm. 
\end{definition}
We note that the \textit{CCNS distance} is best at low values and that we do not define CCNS for regression tasks, since it requires the existence of class labels. 

\subsection{$K$-hop metrics}
Most GAMs only evaluate direct neighbourhoods. However, GNNs can apply the message passing scheme to more hops, including more hops in the node feature embedding. 
We therefore propose to extend homophily and CCNS on unweighted graphs to $k$-hop neighbourhoods. An extension of the metrics on weighted graphs is more challenging, since the edge weights impact the $k$-hop metrics.
The formal definitions for $k$-hop homophily and CCNS for unweighted graphs can be found in the Appendix. 
We here exchange the notion of $\mathcal{N}_v$ with the specific $k$-hop neighbourhood $\mathcal{N}^{(k)}_v$ of interest.

\subsection{Metrics for continuous adjacency matrices}
Several graph learning settings, such as \cite{cosmo2020latent,kazi2022differentiable}, utilise a continuous graph structure.
In order to allow for an evaluation of those graphs, we here define GAMs on the weight matrix $\mathbf{W}$ instead of the binary adjacency matrix $\mathbf{A}$.

\begin{definition}[Homophily for continuous adjacency matrices]
Let $\mathit{G}_w = (\mathit{V}_w, \mathit{E}_w, \mathbf{W})$, be a weighted graph defined as above with a continuous adjacency matrix. Then the $1$-hop node homophily of $\mathit{G}_w$ is defined as follows:

\begin{equation}
    \operatorname{HCont}(G_w, Y) := \frac{1}{|V|} \sum_{v \in V}{ \left(  \frac{\sum_{u \in \mathcal{N}_v | y_u = y_v}{w_{uv}}}{\sum_{u \in \mathcal{N}_v}{w_{uv}}} \right)},
\end{equation}
where $w_{uv}$ is the weight of the edge from $u$ to $v$.
\end{definition}

\begin{definition}[CCNS for continuous adjacency matrices]
Let $\mathit{G}_w = (\mathit{V}_w, \mathit{E}_w, \mathbf{W})$, $C$, $cossim(\cdot, \cdot)$ be defined as above. Then, the CCNS for weighted graphs is defined as follows:
\begin{equation}
    \operatorname{CCNS}_{cont}(c, c') := \frac{1}{\vert \mathcal{V}_c \vert \vert \mathcal{V}_{c'} \vert} \sum_{u \in \mathcal{V}, v \in \mathcal{V'}}{\operatorname{cossim}(d_c(u),d_c(v))},
\end{equation}
\end{definition}
where $d_c(u)$ is the histogram considering the edge weights of the continuous adjacency matrix of the respective classes instead of the count of neighbours. 
The \textit{CCNS distance} for continuous adjacency matrices can be evaluated as above.

\subsection{Homophily for regression}
Homophily is only defined for node classification tasks, which strictly limits its application to a subset of use cases. However, many relevant graph learning tasks perform a downstream node regression, such as age regression \cite{stankeviciute2020population,bintsi2023multimodal}. 
We here define homophily for node regression tasks. Since homophily is a metric ranging from $0$ to $1$, we contain this range for regression tasks by normalising the labels between $0$ and $1$ prior to metric evaluation. We subtract the average node label distance from $1$ to ensure the same range as homophily for classification.

\begin{definition}[Homophily for regression]
    Let $\mathit{G} = (\mathit{V}, \mathit{E})$ and $\mathcal{N}^k_v$ be defined as above and $Y$ be the vector of node labels, which is normalised between $0$ and $1$. Then the $k$-hop homophily for regression is defined as follows:
    \begin{equation}
        \operatorname{HReg}^{(k)}(G, Y) := 1 - \left(  \frac{1}{\vert V \vert} \sum_{v \in V} \left( {\frac{1}{\vert \mathcal{N}^{(k)}_v \vert} \sum_{n \in \mathcal{N}^{(k)}_v}{\left\lVert y_v - y_n \right\rVert_1 }} \right) \right),
    \end{equation}
    where $\lVert \cdot \rVert_1$ indicates the $L_1$ norm.
\end{definition}

\begin{definition}[Homophily for continuous adjacency matrices for regression]
Let $\mathit{G}_w = (\mathit{V}_w, \mathit{E}_w, \mathbf{W})$, Y, and $\mathit{N}_v$ be defined as above and the task be a regression task, then the homophily of $\mathit{G}$ is defined as follows:

\begin{equation}
        \operatorname{HReg}(G, Y) := 1 - \left(  \frac{1}{\vert V \vert} \sum_{v \in V} \left( \frac{ \sum_{n \in \mathcal{N}_v}{w_{nv} \left\lVert y_v - y_n \right\rVert_1 }}{\sum_{n \in \mathcal{N}_v}{w_{nv}}} \right) \right),
\end{equation}
where $w_{nv}$ is the weight of the edge from $n$ to $v$ and $\lVert \cdot \rVert_1$ the $L_1$ norm.
\end{definition}

\subsection{Metric evaluation}
In general, we recommend the evaluation of GAMs separately on the train, validation, and test set. We believe this to be an important evaluation step since the metrics can differ significantly between the different sub-graphs, given that the graph structure in only optimised on the training set. 

\section{Experiments and Results}
We evaluate our metrics on several datasets with different graph learning techniques: We (1) assess benchmark classification datasets using a standard learning pipeline, and (2) medical population graphs for regression and classification that learn the adjacency matrix end-to-end. All experiments are performed in a transductive learning setting using graph convolutional networks (GCNs) \cite{kipf2016semi}. In order to evaluate all introduced GAMs, we specifically perform experiments on two task settings: classification and regression, and under two graph learning settings: one using a discrete adjacency matrix and one using a continuous one.

\subsection{Datasets}

In order to evaluate the above defined GAMs, we perform node-level prediction experiments with GNNs on different datasets.
We evaluate $\{1,2,3\}$-hop homophily and CCNS distance on the benchmark citation datasets Cora, CiteSeer, and PubMed \cite{yang2016revisiting}, Computers and Photos, and Coauthors CS datasets \cite{shchur2018pitfalls}. All of these datasets are classification tasks. We use $k$-layer GCNs and compare performance to a multi-layer perceptron (MLP).

Furthermore, we evaluate the introduced metrics on two different medical population graph datasets, as well as two synthetic datasets. The baseline results for these datasets can be found in Appendix Table \ref{tab:app_baselines}. 
We generate \textbf{synthetic datasets} for classification and regression to analyse the metrics in a controllable setting.
As a real-world medical classification dataset, we use \textbf{TADPOLE} \cite{mueller2005alzheimer}, a neur-imaging dataset which has been frequently used for graph learning on population graphs \cite{parisot2017spectral,cosmo2020latent,kazi2022differentiable}.
For a regression population graph, we perform brain age prediction on $6\,406$ subjects of the UK BioBank \cite{ukbb} (\textbf{UKBB}). We use $22$ clinical and $68$ imaging features extracted from the subjects' magnetic resonance imaging (MRI) brain scans, following the approach in \cite{cole2020multimodality}. In both medical population graphs, each subject is represented by one node and similar subjects are either connected following the $k$-nearest neighbours approach, like in \cite{kazi2022differentiable} or starting without any edges.

\subsection{GNN Training}
Prior to this work, the homophily metric has only existed for an evaluation on discrete adjacency matrices. In this work, we extend this metric to continuous adjacency matrices. In order to evaluate the metrics for both, discrete and continuous adjacency matrices, we use two different graph learning methods: (a) \textit{dDGM} and (b) \textit{cDGM} from \cite{kazi2022differentiable}. DGM stands for \say{differentiable graph module}, referring to the fact that both methods learn the adjacency matrix in an end-to-end manner. cDGM hereby uses a continuous adjacency matrix, allowing us to evaluate the metrics introduced specifically for this setting. dDGM uses a discrete adjacency matrix by sampling the edges using the Gumbel-Top-K trick \cite{jiang2019semi}. Both methods are similar in terms of model training and performance, allowing us to compare the newly introduced metrics to the existing homophily metric in the dDGM setting.

\subsection{Results}
\subsubsection{(1) Benchmark classification datasets}
The results on the benchmark datasets are summarised in Table \ref{tab:gna_benchmark}.  We can see that the $k$-hop metric values can differ greatly between the different hops for some datasets, while staying more constant for others. This gives an interesting insight into the graph structure over several hops. 
We believe an evaluation of neighbourhoods in graph learning to be more insightful if the number of hops in the GNN matches the number of hops considered in the graph metric. 
Interestingly, performance of $k$-hop GCNs did not align with the $k$-hop metric values on the specific datasets. 
We summarise these results in Appendix Table \ref{tab:app_benchmarks}. One possible reason for this might be that, e.g., the $3$-hop metrics assess the $1$, $2$, and $3$-hop neighbourhood at once, not just the outer ring of neighbours. Another reason for this discrepancy might be that homophily and CCNS do not perfectly predict GNN performance. Furthermore, different graph convolutions have shown to be affected differently by low-homophily graphs \cite{zhu2020beyond}. We believe this to be an interesting direction to further investigate GAMs for GNNs.

\begin{table}[t]
\centering
\addtolength{\tabcolsep}{3pt}
\scriptsize
\caption{$K$-hop graph metrics of benchmark node classification datasets. Cl.: number of classes, Nodes: number of nodes}
\begin{tabular}{lrrcccccccc}
\toprule[0.8pt]
\multirow{2}{*}{\textbf{Dataset}} &  \multirow{2}{*}{\textbf{Nodes}} & \multirow{2}{*}{\textbf{Cl.}} & \multicolumn{3}{c}{\textbf{Node homophily  $\uparrow$}} &  \multicolumn{3}{c}{$D_{\operatorname{CCNS}}$ $\downarrow$} \\
& & & \textbf{$1$-hop} &\textbf{$2$-hop} & \textbf{$3$-hop} & \textbf{$1$-hop} &\textbf{$2$-hop} & \textbf{$3$-hop}  \\
\toprule[0.8pt]
Cora   & 1,433 & 7 &  0.825 {\tiny $\pm$ 0.29} & 0.775  {\tiny $\pm$ 0.26} & 0.663 {\tiny $\pm$ 0.29} & 0.075 & 0.138 & 0.229 \\
CiteSeer  & 3,703 & 6 & 0.706 {\tiny $\pm$ 0.40} & 0.754 {\tiny $\pm$ 0.28} & 0.712 {\tiny $\pm$ 0.29}  &0.124 & 0.166 & 0.196 \\ 
PubMed & 19,717 & 3 & 0.792 {\tiny $\pm$ 0.35} & 0.761 {\tiny $\pm$ 0.26} & 0.687 {\tiny $\pm$ 0.26} & 0.173 & 0.281 & 0.363 \\
\midrule
Computers &13,752 & 10  & 0.785 {\tiny $\pm$ 0.26} & 0.569 {\tiny $\pm$ 0.27} & 0.303 {\tiny $\pm$ 0.20} & 0.080 & 0.275 & 0.697 \\
Photo & 7,650 & 8 & 0.837 {\tiny $\pm$ 0.25} & 0.660 {\tiny $\pm$ 0.30} & 0.447 {\tiny $\pm$ 0.28} & 0.072 & 0.210 & 0.429 \\
\midrule
Coauthor CS & $18,333$ & $15$ & 0.832 {\tiny $\pm$ 0.24} & 0.698 {\tiny $\pm$ 0.25} & 0.520 {\tiny $\pm$ 0.25} & 0.043 & 0.110 & 0.237 \\
\bottomrule
\end{tabular}
\label{tab:gna_benchmark}
\end{table}

\subsubsection{(2) Population graph experiments}
Table \ref{tab:pop_graph_res} shows the dDGM and cDGM results of the population graph datasets.
We can see that in some settings, such as the classification tasks on the synthetic dataset using dDGM, the homophily varies greatly between train and test set. 
This can be an indication for over-fitting on the training set, since the graph structure is optimised for the training nodes only and might not generalise well to the whole graph. 

\begin{figure}[t]
    \centering
    \includegraphics[width=\textwidth]{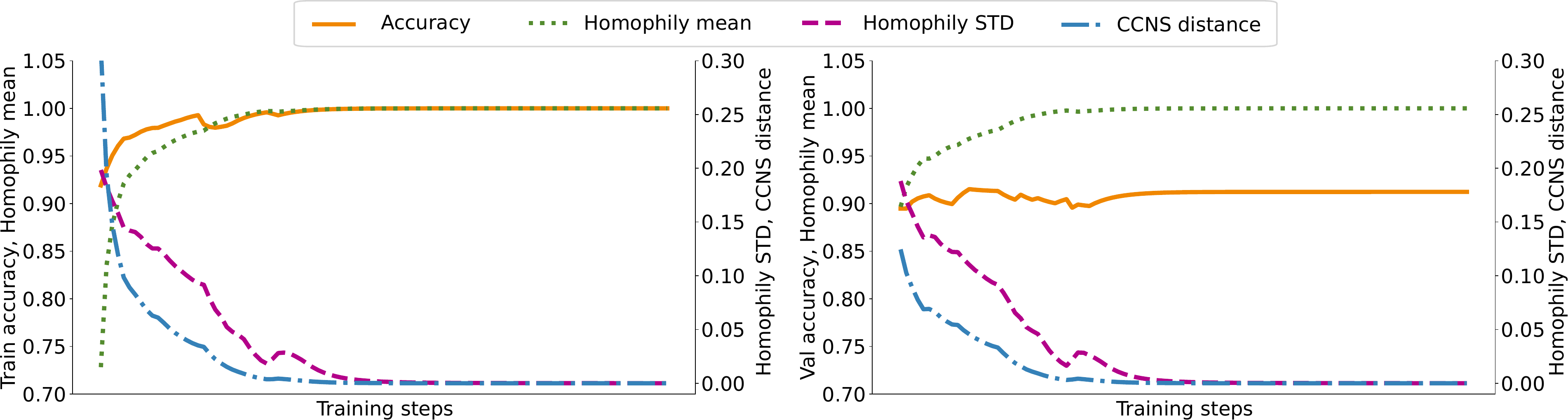}
    \caption{Development of graph metrics on TADPOLE over training using \textbf{cDGM}; left: train set; right: validation set}
    \label{fig:res_tadpole}
\end{figure}

\begin{figure}[t]
    \centering
    \includegraphics[width=\textwidth]{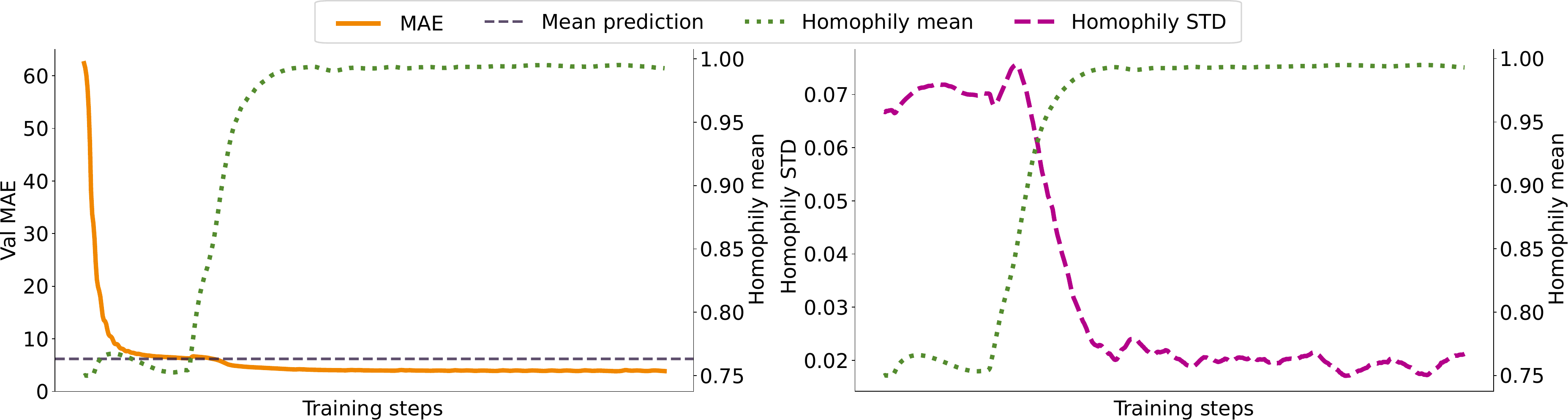}
    \caption{Development of metrics on UKBB dataset using \textbf{cDGM} on validation set}
    \label{fig:ukbb_cdgm}
\end{figure}

\begin{table*}[t]
\centering
\scriptsize
\addtolength{\tabcolsep}{3pt}
\caption{cDGM and dDGM results on the population graph datasets. We report the test scores averaged over $5$ random seeds and $1$-hop homophily and CCNS distance of one final model each. We do not report CCNS distance on regression datasets, since it is not defined for regression tasks.}
\begin{tabular}{lllccccc}
    \toprule
    \multirow{2}{*}{\textbf{Method}} &
    \multirow{2}{*}{\textbf{Dataset}} &
    \multirow{2}{*}{\textbf{Task}} &
    \multirow{2}{*}{\textbf{Test score}}&
    \multicolumn{2}{c}{\textbf{1-hop node homophily $\uparrow$}}& 
    \multicolumn{2}{c}{\textbf{$1$-hop} $D_{\operatorname{CCNS}}$ $\downarrow$} \\
     &&&& \textbf{train} & \textbf{test} & \textbf{train} & \textbf{test}  \\
    \toprule
     \textbf{cDGM} & Synthetic 1k  & c & 0.7900 {\tiny $\pm$ 0.08} & 1.0000 {\tiny $\pm$ 0.00} & 1.0000 {\tiny $\pm$ 0.00} & 0.0000 & 0.0000 \\ 
     & & r & 0.0112 {\tiny $\pm$ 0.01} & 0.9993 {\tiny $\pm$ 0.00} & 0.9991 {\tiny $\pm$ 0.00} & - & - \\ 
     \cmidrule{2-8}
     & Synthetic 2k & c & 0.8620 {\tiny $\pm$ 0.03} & 1.0000  {\tiny $\pm$ 0.00} & 1.0000 {\tiny $\pm$ 0.00} & 0.0000 & 0.0000 \\
     && r & 0.0173 {\tiny $\pm$ 0.00} & 0.8787 {\tiny $\pm$ 0.06} & 0.8828 {\tiny $\pm$ 0.05} & - & - \\
     \cmidrule{2-8}
    & Tadpole  & c & 0.9333 {\tiny $\pm$ 0.01} & 1.0000 {\tiny $\pm$ 0.00} & 0.9781  {\tiny $\pm$ 0.09} & 0.0000 & 0.0314 \\
     \cmidrule{2-8}
    & UKBB  & r & 4.0775 {\tiny $\pm$ 0.23} & 0.8310 {\tiny $\pm$ 0.06} & 0.8306 {\tiny $\pm$ 0.07} & - & - \\ 
    \midrule
    \midrule
    \textbf{dDGM} & Synthetic 1k & c & 0.8080 {\tiny $\pm$ 0.04} & 0.6250 {\tiny $\pm$ 0.42} & 0.1150 {\tiny $\pm$ 0.32} & 0.4483 &  0.4577 \\
    & & r & $0.0262$ {\tiny $\pm$ $0.00$} &  {0.7865 \tiny $\pm$ 016} &  {0.8472 \tiny $\pm$ 0.15} & - & - \\
     \cmidrule{2-8}
    & Synthetic 2k & c &  0.7170 {\tiny $\pm$ 0.06} & 0.6884 {\tiny $\pm$ 0.40} & 0.0950 {\tiny  $\pm$ 0.29 } & 0.4115 & 0.4171 \\
    & & r & 0.0119 {\tiny $\pm$ 0.00} &  0.8347 {\tiny $\pm$ 0.13} & 0.8295 {\tiny $\pm$ 0.13 } & - & - \\
 \cmidrule{2-8}
& Tadpole & c & 0.9614 {\tiny $\pm$  0.01} & 0.9297 {\tiny $\pm$ 0.18} & 0.8801 {\tiny $\pm$ 0.31} & 0.1045 & 0.0546 \\
 \cmidrule{2-8}
& UKBB & r & 3.9067 {\tiny $\pm$ 0.04} & 0.8941 {\tiny $\pm$ 0.13} & 0.9114 {\tiny $\pm$ 0.12} & - & -\\
\bottomrule
\end{tabular}
\label{tab:pop_graph_res}
\end{table*}

Since we here use graph learning methods which adapt the graph structure during model training, also the graph metrics change over training. Figure \ref{fig:res_tadpole} shows the development of the accuracy as well as the mean and standard deviation of the $1$-hop homophily and CCNS distance, evaluated on the train (left) and validation set (right). 
We can see that for both sets, the homophily increases with the accuracy, while the standard deviation (STD) of the homophily decreases and the CCNS distance decreases with increasing performance. However, the GAMs align more accurately with the training accuracy (left), showing that the method optimised the graph structure on the training set. The validation accuracy does not improve much in this example, while the validation GAMs still converge similarily to the ones evaluated on the train set (left).
Figure \ref{fig:ukbb_cdgm} shows the mean (left) and STD (right) of the validation regression homophily $\operatorname{HReg}$ on the UKBB dataset with continuous adjacency matrices (using cDGM) and the corresponding change in validation mean absolute error (MAE). 
Again, homophily raises when the validation MAE decreases and the STD of the homophily decreases in parallel. On the left, the dotted grey line indicates the MAE of a mean prediction on the dataset. 
We can see that the mean regression homophily $\operatorname{HReg}$ raises once the validation MAE drops below the error of a mean prediction. We here only visualise a subset of all performed experiments, but we observe the same trends for all settings. From these experiments we conclude that the here introduced GAMs show strong correlation with model performance and can be used to assess generated graph structures that are used for graph deep learning. 

\section{Conclusion and Future Work}
In this work, we extended two frequently used graph assessment metrics (GAMs) for graph deep learning, that allow to evaluate the graph structure in regression tasks and continuous adjacency matrices. 
For datasets that do not come with a pre-defined graph structure, like population graphs, the assessment of the graph structure is crucial for quality checks on the learning pipeline.
Node homophily and cross-class neighbourhood similarity (CCNS) are commonly used GAMs that allow to evaluate how similar the neighbourhoods in a graph are. 
However, these metrics are only defined for discrete adjacency matrices and classification tasks. 
This only covers a small portion of graph deep learning tasks. Several graph learning tasks target node regression \cite{stankeviciute2020population,bintsi2023multimodal,math10050786}. 
Furthermore, recent graph learning methods have shown that an end-to-end learning of the adjacency matrix is beneficial over statically creating the graph structure prior to learning \cite{kazi2022differentiable}. 
These methods do not operate on a static binary adjacency matrix, but use weighted continuous graphs, which is not considered by  most current GAMs.
In order to overcome these limitations, we extend the definition of node homophily to regression tasks and both node homophily and CCNS to continuous adjacency matrices. 
We formulate these metrics and evaluate them on different synthetic and real world medical datasets and show their strong correlation with model performance. 
We believe these metrics to be essential tools for investigating the performance of GNNs, especially in the setting of population graphs or similar settings that require explicit graph construction. 

Our definition of the CCNS distance $D_{\operatorname{CCNS}}$ uses the $L_1$-norm to determine the distance between the node labels in order to weight each inter-class-connection equally. However, the $L_1$-norm is only one of many norms that could be used here. Given the strong correlation of our definition of $D_{\operatorname{CCNS}}$, we show the the usage of the $L_1$-norm is a sensible choice. 
We also see an extension of the metrics for weighted graphs to multiple hops as promising next steps towards better graph assessment for GNNs.

There exist additional GAMs, such as normalised total variation and normalised smoothness value \cite{lu2022nagnn}, neighbourhood entropy and centre-neighbour similarity \cite{xie2020gnns}, and aggregations similarity score and diversification distinguishability \cite{elam2021human} that have been shown to correlate with GNN performance. 
An extension of these metrics to regression tasks and weighted graphs would be interesting to investigate in future works. 
All implementations of the here introduced metrics are differentiable. 
This allows for a seamless integration in the learning pipeline, e.g. as loss components, which could be a highly promising application to improve GNN performance by optimising for specific graph properties.


\bibliographystyle{splncs04}
\bibliography{literature}

\newpage
\appendix

\section{Further Information on Extended Graph Assessment Metrics}
\subsection{$K$-hop metrics}
We here formally define $k$-hop node homophily and $k$-hop CCNs.

\begin{definition}[$k$-hop node homophily]
  A graph $\mathit{G} := (\mathit{V}, \mathit{E})$ with the set of node labels $\mathit{Y} := \{y_u; u \in \mathit{V}\}$ has the following $k$-hop node homophily:

\begin{equation}
  h^{(k)}(\mathit{G}, \mathit{Y}) := \frac{1}{\vert \mathit{V} \vert} \sum_{v \in \mathit{V}}\frac{\left\vert \{ u | u \in \mathcal{N}^{(k)}_v, y_u = y_v \} \right\vert}{\vert \mathcal{N}^{(k)}_v \vert},
  \label{eq:khophomophily}
\end{equation} 
where $\mathcal{N}^{(k)}_v$ is the set of nodes in the $k$-hop neighbourhood of $v$.
\end{definition}

\begin{definition}[$k$-hop CCNS]
  A graph $\mathit{G} := (\mathit{V}, \mathit{E})$ with the set of node labels $C$ has the following $k$-hop CCNS for two classes $c$ and $c'$:

\begin{equation}
    \operatorname{CCNS}(c, c') = \frac{1}{\vert \mathcal{V}_c \vert \vert \mathcal{V}_{c'} \vert} \sum_{u \in \mathcal{V}, v \in \mathcal{V'}}{\operatorname{cossim}(d^{(k)}(u),d^{(k)}(v))},
\end{equation}
where $d^{(k)}(v)$ indicates the empirical histogram of the labels of the $k$-hop neighbours of node $v$ and $\operatorname{cossim}(\cdot,\cdot)$ the cosine similarity. 
\end{definition}

Table \ref{tab:app_benchmarks} summarises model performances of $\{1,2,3\}$-hop GCNs on the different benchmark datasets and the corresponing MLP performance on the node features only. We can see that even though $3$-hop homophily  of the datasets Computers and Photo is very low, the GCNs with $3$ hops perform best on these datasets. This does not align with our initial intuition about these metrics and we believe this finding to be interesting to investigate further.

\begin{table*}[!ht]
\centering
\addtolength{\tabcolsep}{1pt}
\scriptsize
\caption{Graph metrics of benchmark node classification datasets with corresponding performances of an MLP and 1,2, and 3-hop GCNs, reported in accuracy in \%. Nodes: number of nodes, Cl.: number of classes in the dataset.} 
\begin{tabular}{lrrccccccccc}
\toprule[0.8pt]
\multirow{2}{*}{\textbf{Dataset}} &  \multirow{2}{*}{\textbf{Nodes}} & \multirow{2}{*}{\textbf{Cl.}} & \multicolumn{3}{c}{\textbf{Node homophily}}  & \multirow{2}{*}{\textbf{MLP}}&  \multicolumn{3}{c}{\textbf{GCN}} \\
& & & \textbf{1-hop} &\textbf{2-hop} & \textbf{3-hop} & & \textbf{1-hop} & \textbf{2-hop} & \textbf{3-hop}\\
\toprule[0.8pt]
Cora & $1\,433$ & 7 &  0.825 {\tiny $\pm$ 0.29} & 0.775  {\tiny $\pm$ 0.26} & 0.663 {\tiny $\pm$ 0.29} & 60.41 & 76.33 & \textbf{81.70} & 78.90 \\
Citeseer  & $3\,703$ & 6 & 0.706 {\tiny $\pm$ 0.40} & 0.754 {\tiny $\pm$ 0.28} & 0.712 {\tiny $\pm$ 0.29} & 61.19 & 71.20 & \textbf{72.10} & 67.10  \\ 
Pubmed  & $19\,717$ & 3 & 0.792 {\tiny $\pm$ 0.35} & 0.761 {\tiny $\pm$ 0.26} & 0.687 {\tiny $\pm$ 0.26} & 74.00 & 76.60 & \textbf{79.10} & 77.70  \\
\midrule
Computers & $13\,752$ & 10  & 0.785 {\tiny $\pm$ 0.26} & 0.569 {\tiny $\pm$ 0.27} & 0.303 {\tiny $\pm$ 0.20} & 79.35 & 39.27 & 67.56 & \textbf{83.13} \\
Photo & $7\,650$ & 8 & 0.837 {\tiny $\pm$ 0.25} & 0.660 {\tiny $\pm$ 0.30} & 0.447 {\tiny $\pm$ 0.28} & 82.09 & 48.10 & 82.88 & \textbf{88.37} \\
\midrule
Coauthor CS & $18\,333$ & $15$ & 0.832 {\tiny $\pm$ 0.24} & 0.698 {\tiny $\pm$ 0.25} & 0.520 {\tiny $\pm$ 0.25} & 88.93 & \textbf{93.13} & 89.31 & 92.09 \\
\bottomrule
\end{tabular}
\label{tab:app_benchmarks}
\end{table*}

\subsection{Node-wise metrics}
The $k$-hop homophily for regression can also defined for every node individually and then combined in the full homophily over the entire graph as defined in the main part of this work.

\begin{definition}[Homophily for regression]
Let $\mathit{G} = (\mathit{V}, \mathit{E})$ and $\mathcal{N}^k_v$ be defined as above and $Y$ be the vector of node labels, which is normalised between $0$ and $1$. Then the $k$-hop homophily of a node $v \in \mathit{V}_c$ in a node regression task is defined as the mean label distance between the node $v$ and all it's neighbours.
\begin{equation}
    \operatorname{HReg}^{(k)}_v := 1 - \left( \frac{1}{\vert \mathcal{N}^{(k)}_v \vert} \sum_{n \in \mathcal{N}^{(k)}_v}{\left\lVert y_v - y_n \right\rVert } \right),
\end{equation}
where $\vert \cdot \vert$ is the cardinality of a set and $\lVert x \rVert$ the absolute value of x.
\end{definition}
The $k$-hop homophily for regression of the whole graph $\mathit{G}$ can then be extracted as follows:
\begin{align}
    \operatorname{HReg}^{(k)}_G &:= 1 - \left( \frac{1}{\vert V \vert} \sum_{v \in V}{\operatorname{HReg}^{(k)}_v} \right) \nonumber \\
    &= 1 - \left(  \frac{1}{\vert V \vert} \sum_{v \in V} \left( {\frac{1}{\vert \mathcal{N}^{(k)}_v \vert} \sum_{n \in \mathcal{N}^{(k)}_v}{\left\lVert y_v - y_n \right\rVert }} \right) \right).
\end{align}

\section{Experiments}
In this section we give more details on training parameters and setups of the experiments performed in this work.
\subsection{Synthetic dataset}
The synthetic datasets are generated using \textit{sklearn} \cite{sklearn_api}. Each dataset consist of either $1\,000$ or $2\,000$ nodes, with $50$ node features of which $5$ are informative. For all experiments on the synthetic dataset we utilise early stopping and the initial graph structure is generated using the $k$-nearest neighbours approach with $5$ neighbours and the Euclidean distance.

\subsection{TADPOLE dataset}
We use the same TADPOLE dataset as in \cite{kazi2022differentiable}, which consists of $564$ subjects. The task is the classification of Alzheimer's disease, mild cognitive impairment and control normal. For all experiments on the TADPOLE dataset, we use early stopping and generate the initial graph structure using the $k$-nearest neighbours approach and the Euclidean distance.

\subsection{UKBB dataset}
The dDGM experiments on the UKBB dataset are performed using no initial graph structure, since this resulted in better model performance. For the cDGM experiments we use the $k$-nearest neighbours approach with $x$ neighbours. We utilised no early stopping and the Euclidean distance for the graph construction for the cDGM experiments.

\subsection{Baseline results}
Table \ref{tab:app_baselines} summarises the baseline results on the population graph datasets using a random forest and the implementation from sklearn \cite{sklearn_api}.

\begin{table*}[ht]
\centering
\scriptsize
\addtolength{\tabcolsep}{5pt}
\caption{Baseline results using random forests on the different datasets. For classification tasks, we report accuracy in \% and for regression MAE. We report the mean and standard deviation of a $5$-fold cross validation.}
\begin{tabular}{lllll}
\toprule
 \textbf{Dataset} & \textbf{Nr. nodes} & \textbf{Task} &  \textbf{Test Score} \\
 \midrule
 Synthetic & $1000$ & Binary classification & 78.00 $\pm$ 0.07 \\
 & & Regression & 0.0529 $\pm$ 0.01 \\
 \cmidrule{2-4}
 & $2000$ & Binary classification & 88.10 $\pm$ 0.02 \\
 & & Regression & 0.0081 $\pm$ 0.00 \\
 \midrule
 Tadpole & $564$ & Classification & 94.15 $\pm$ 0.01 \\
 \midrule
 UKBB & $6406$ & Regression & 4.2644 $\pm$ 0.05 \\
\bottomrule
\end{tabular}
\label{tab:app_baselines}
\end{table*}

\end{document}